\documentclass{aastex}
\usepackage{amsmath}
\usepackage{amssymb}

\begin{document}

\title{PAMELA's measurements of geomagnetically trapped and albedo protons}
\shorttitle{PAMELA's measurements of geomagnetically trapped and albedo protons}
\shortauthors{Bruno et al.}

\author{
A.~Bruno$^{1,*}$,
O.~Adriani$^{2,3}$,
G.~C.~Barbarino$^{4,5}$,
G.~A.~Bazilevskaya$^{6}$,
R.~Bellotti$^{1,7}$,
M.~Boezio$^{8}$,
E.~A.~Bogomolov$^{9}$,
M.~Bongi$^{2,3}$,
V.~Bonvicini$^{8}$,
S.~Bottai$^{3}$,
U.~Bravar$^{10}$,
F.~Cafagna$^{7}$,
D.~Campana$^{5}$,
R.~Carbone$^{8}$,
P.~Carlson$^{11}$,
M.~Casolino$^{12,13}$,
G.~Castellini$^{14}$,
E.~C.~Christian$^{15}$,
C.~De~Donato$^{12,17}$,
G.~A.~de~Nolfo$^{15}$,
C.~De~Santis$^{12,17}$,
N.~De~Simone$^{12}$,
V.~Di~Felice$^{12,18}$,
V.~Formato$^{8,19}$,
A.~M.~Galper$^{16}$,
A.~V.~Karelin$^{16}$,
S.~V.~Koldashov$^{16}$,
S.~Koldobskiy$^{16}$,
S.~Y.~Krutkov$^{9}$,
A.~N.~Kvashnin$^{6}$,
M.~Lee$^{10}$,
A.~Leonov$^{16}$,
V.~Malakhov$^{16}$,
L.~Marcelli$^{12,17}$,
M.~Martucci$^{17,20}$,
A.~G.~Mayorov$^{16}$,
W.~Menn$^{21}$,
M.~Merg\`e$^{12,17}$,
V.~V.~Mikhailov$^{16}$,
E.~Mocchiutti$^{8}$,
A.~Monaco$^{1,7}$,
N.~Mori$^{2,3}$,
R.~Munini$^{8,19}$,
G.~Osteria$^{5}$,
F.~Palma$^{12,17}$,
B.~Panico$^{5}$,
P.~Papini$^{3}$,
M.~Pearce$^{11}$,
P.~Picozza$^{12,17}$,
M.~Ricci$^{20}$,
S.~B.~Ricciarini$^{3,14}$,
J.~M.~Ryan$^{10}$,
R.~Sarkar$^{22,23}$,
V.~Scotti$^{4,5}$,
M.~Simon$^{21}$,
R.~Sparvoli$^{12,17}$,
P.~Spillantini$^{2,3}$,
S.~Stochaj$^{24}$,
Y.~I.~Stozhkov$^{6}$,
A.~Vacchi$^{8}$,
E.~Vannuccini$^{3}$,
G.~I.~Vasilyev$^{9}$,
S.~A.~Voronov$^{16}$,
Y.~T.~Yurkin$^{16}$,
G.~Zampa$^{8}$,
N.~Zampa$^{8}$,
and V.~G.~Zverev$^{16}$.
}

\affil{$^{1}$ Department of Physics, University of Bari ``Aldo Moro'', I-70126 Bari, Italy.}
\affil{$^{2}$ Department of Physics and Astronomy, University of Florence, I-50019 Sesto Fiorentino, Florence, Italy.}
\affil{$^{3}$ INFN, Sezione di Florence, I-50019 Sesto Fiorentino, Florence, Italy.}
\affil{$^{4}$ Department of Physics, University of Naples ``Federico II'', I-80126 Naples, Italy.}
\affil{$^{5}$ INFN, Sezione di Naples, I-80126 Naples, Italy.}
\affil{$^{6}$ Lebedev Physical Institute, RU-119991 Moscow, Russia.}
\affil{$^{7}$ INFN, Sezione di Bari, I-70126 Bari, Italy.}
\affil{$^{8}$ INFN, Sezione di Trieste, I-34149 Trieste, Italy.}
\affil{$^{9}$ Ioffe Physical Technical Institute, RU-194021 St. Petersburg, Russia.}
\affil{$^{10}$ Space Science Center, University of New Hampshire, Durham, NH, USA.}
\affil{$^{11}$ KTH, Department of Physics, and the Oskar Klein Centre for Cosmoparticle Physics, AlbaNova University Centre, SE-10691 Stockholm, Sweden.}
\affil{$^{12}$ INFN, Sezione di Rome ``Tor Vergata'', I-00133 Rome, Italy.}
\affil{$^{13}$ RIKEN, Advanced Science Institute, Wako-shi, Saitama, Japan.}
\affil{$^{14}$ IFAC, I-50019 Sesto Fiorentino, Florence, Italy.}
\affil{$^{15}$ Heliophysics Division, NASA Goddard Space Flight Ctr, Greenbelt, MD, USA.}
\affil{$^{16}$ National Research Nuclear University MEPhI, RU-115409 Moscow, Russia.}
\affil{$^{17}$ Department of Physics, University of Rome ``Tor Vergata'', I-00133 Rome, Italy.}
\affil{$^{18}$ Agenzia Spaziale Italiana (ASI) Science Data Center, %Via del Politecnico snc,
I-00133 Rome, Italy.}
\affil{$^{19}$ Department of Physics, University of Trieste, I-34147 Trieste, Italy.}
\affil{$^{20}$ INFN, Laboratori Nazionali di Frascati, %Via Enrico Fermi 40,
I-00044 Frascati, Italy.}
\affil{$^{21}$ Department of Physics, Universit\"{a}t Siegen, D-57068 Siegen, Germany.}
\affil{$^{22}$ Indian Centre for Space Physics, 43 Chalantika, %Garia Station Road,
Kolkata 700084, West Bengal, India.}
\affil{$^{23}$ Previously at INFN, Sezione di Trieste, I-34149 Trieste, Italy. }
\affil{$^{24}$ Electrical and Computer Engineering, New Mexico State University, Las Cruces, NM, USA.}

\altaffiltext{*}{Corresponding author. E-mail address: alessandro.bruno@ba.infn.it.}

\begin{abstract}
Data from the PAMELA satellite experiment were used to perform a detailed measurement of
under-cutoff protons at low Earth orbits. On the basis of a trajectory tracing approach using
a realistic description of the magnetosphere, protons were classified into geomagnetically
trapped and
re-entrant
albedo. The former
include
\emph{stably-trapped} protons in the South Atlantic Anomaly,
which were analyzed in the framework of the adiabatic theory, investigating energy spectra,
spatial and angular distributions; 
results
were compared with
the predictions of the AP8 and the PSB97 empirical trapped models.
The albedo protons were classified into
\emph{quasi-trapped}, concentrating in the magnetic equatorial region, and \emph{un-trapped}, spreading
over all latitudes and including both short-lived (\emph{precipitating}) and long-lived (\emph{pseudo-trapped})
components. Features of the penumbra region around the geomagnetic cutoff were investigated
as well.
PAMELA
observations
significantly improve the characterization of the high energy proton
populations in near Earth orbits.
\end{abstract}

%\FullConference{The 34th International Cosmic Ray Conference,\\
%		30 July- 6 August, 2015\\
%		The Hague, The Netherlands}

\section{Introduction}\label{Introduction}
The radiation environment
in Earth's vicinity constitutes a well-known hazard for the space missions. Major sources include large solar particle events and the Van Allen belts, consisting of intense fluxes of energetic charged particles experiencing long-term magnetic trapping.
Speci\-fi\-cally, the inner belt is mainly populated by protons, mostly originated by the decay of albedo neutrons according to the CRAND mechanism \citep{Walt}. A standard description of such an environment is provided by the AP8 empirical model \citep{AP8}, based on data from 
satellite experiments in the 1960s and early 1970s. Recently, significant improvements \citep{CRRESPRO,Huston,PSB97,TPM1} have been made thanks to the data from new spacecrafts \citep{CRRES,CRRES2,Looper96,Looper98,Huston96}. Nevertheless,
the modeling of the 
trapped environment is still incomplete, with
largest uncertainties affecting the high energy 
fluxes in the inner zone and the South Atlantic Anomaly (SAA), where the inner belt makes its closest approach to the Earth\footnote{The SAA is a consequence of the tilt ($\sim$10 deg) between the magnetic dipole axis of the Earth and its rotational axis, and of the offset ($\sim$500 km) between the dipole and the Earth centers.}. 

In addition, the 
magnetospheric radiation
includes populations of albedo protons, originated by the collisions of Cosmic-Rays (CRs) from interplanetary space on the 
atmosphere \citep{Treiman}.
A quasi-trapped component concentrates in the 
equatorial region 
and presents features similar to those of radiation belt
protons,
but 
with limited lifetimes and much less intense fluxes \citep{Moritz,Hovestadt,AMS01}.
An un-trapped component spreads over all latitudes \citep{AMS01b,Bidoli} including the
``penumbra'' region around the geomagnetic cutoff, where particles of both cosmic and atmospheric origin are present \citep{Cooke}.

New accurate measurements of the CR radiation at low Earth orbits have been performed by the PAMELA experiment \citep{PHYSICSREPORTS}.
This paper reports the 
observations of the geomagnetically trapped 
and re-entrant albedo 
protons. 

\section{Data analysis}\label{Data analysis}
PAMELA is a space-based experiment designed for a precise measurement of charged CRs in the energy
range from some tens of MeV up to several hundreds of GeV. 
The Resurs-DK1 satellite, which hosts the apparatus, has a semi-polar (70 deg inclination) and elliptical (350$\div$610 km altitude) orbit.
The spacecraft is 3-axis stabilized; its orientation is calculated by an onboard processor with an ac\-cu\-ra\-cy better than 1 deg.
Particle directions are measured with a high angular resolution ($<$ 2 deg).
Details about apparatus performance, proton selection, detector efficiencies and experimental uncertainties can be found elsewhere (see e.g. \citet{SOLARMOD}).
The data set analyzed
in this work includes protons collected 
by PAMELA
between 2006 July and 2009 September. 

\subsection{Particle classification}\label{Particle classification}
Trajectories of all detected down-going protons were reconstructed in the Earth's magnetosphere using a tracing program based on numerical integration methods \citep{TJPROG,SMART}, and implementing the IGRF11 \citep{IGRF11} and the TS05 \citep{TS05} as internal and external geomagnetic models.
Trajectories were propagated back and forth from the measurement location, and
traced until: they
reached the model magnetosphere boundaries (\emph{galactic} protons); or they intersected the absorbing atmosphere limit, which was assumed at an altitude\footnote{Such a value approximately corresponds to the mean production altitude for albedo protons.}
of 40 km (\emph{re-entrant albedo} protons); or they performed more than $10^{6}/R^{2}$ steps\footnote{Since the program uses a dynamic variable step length, which is of the order of 1\% of a particle gyro-distance in the magnetic field,
such a criterion ensures that at least 4 drift cycles around the Earth were performed.}, where $R$ is the particle rigidity 
in GV, for both propagation directions (\emph{geomagnetically trapped} protons).
Trapped trajectories were verified to fulfil the adiabatic conditions \citep{PAMTRAPPED}, in particular the hierarchy of temporal scales:
$\omega_{gyro} \gg \omega_{bounce} \gg \omega_{drift}$,
where $\omega_{gyro}$, $\omega_{bounce}$ and $\omega_{drift}$ are the frequencies associated with 
gyration, bouncing and drift motions.

Albedo protons were classified into \emph{quasi-trapped} and \emph{un-trapped}. The former have trajectories similar to those of stably-trapped, 
but are originated and re-absorbed by the atmosphere during a time larger than a bounce period (up to several tens of s). The latter 
include both a short-lived component of
protons
\emph{precipitating} into the atmosphere within a bounce period ($\lesssim$ 1s), and a long-lived (\emph{pseudo-trapped}) component with rigidities near the geomagnetic cutoff (penumbra region), 
cha\-ra\-cterized by a chaotic motion (non-adiabatic trajectories).
Further details, including distributions of lifetimes and production/absorption points on the atmosphere, can be found in \citet{ALBEDO}.

\subsection{Flux calculation}\label{Flux calculation}
Proton fluxes were derived by assuming an isotropic flux distribution in all the explored regions except the SAA. In this case, fluxes are significantly anisotropic due to the interaction with the Earth's atmosphere, and thus the \emph{gathering power} of the apparatus \citep{Sullivan} depends on the spacecraft orientation with respect to the geomagnetic field. Consequently, a PAMELA \emph{effective area} (cm$^{2}$) was evaluated as a function of particle energy $E$, local pitch angle $\alpha$ and satellite orientation $\Psi$:
\begin{equation}\label{area_formula}
H(E,\alpha,\Psi)=\frac{  sin\alpha}{2\pi}\int_{0}^{2\pi} d\beta \left[ A(E,\theta,\phi) \cdot cos\theta \right],
\end{equation}
where $\beta$ is the gyro-phase angle, $\theta$=$\theta(\alpha,\beta,\Psi)$ and $\phi$=$\phi(\alpha,\beta,\Psi)$ are respectively the zenith and the azimuth angle describing particle direction in the PAMELA frame\footnote{The PAMELA frame has the origin in the center of the spectrometer cavity; the Z axis is directed along the main axis of the apparatus, toward incoming particles; the Y axis is directed opposite to the main direction of the magnetic field inside the spectrometer; the X axis completes a right-handed system.}, and $A(E,\theta,\phi)$ is the apparatus response function.
The effective area was evaluated with accurate Monte Carlo simulations based on integration methods \citep{Sullivan}.
Finally, in order to account for effects due to the large particle gyro-radius (up to several hundreds of km), trapped fluxes were evaluated by shifting measured protons ($L$, $B$, $B_{eq}$) to corresponding guiding center positions. 
Further details can be found in \citet{PAMTRAPPED}.

\begin{figure}[!t]
\centering
\includegraphics[width=5.5in]{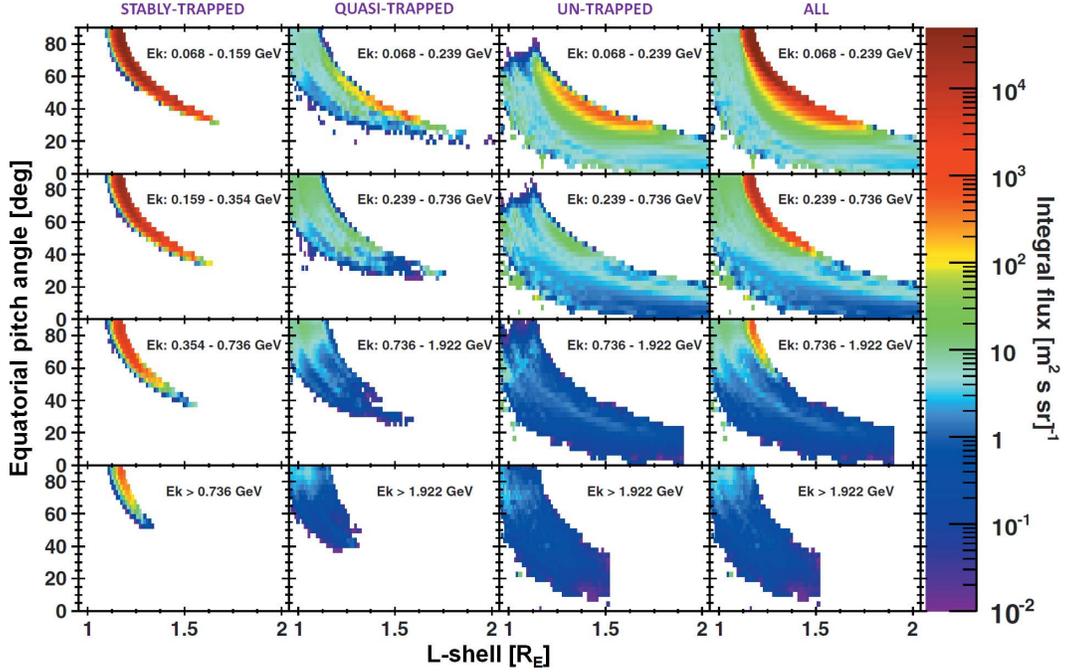}
\caption{Proton integral fluxes ($m^{-2} s^{-1} sr^{-1}$) as a function of equatorial pitch angle $\alpha_{eq}$ and McIlwain's $L$-shell, for different kinetic energy bins (see the labels). Results for the various components are reported (from left to right): stably-trapped, quasi-trapped, un-trapped and the total sample. See the text for details.}
\label{Fig1}
\end{figure}

\section{Results}\label{Results}
Figure \ref{Fig1} shows the fluxes of under-cutoff protons
as a function of equatorial pitch angle $\alpha_{eq}$ and McIlwain's $L$-shell, integrated over different kinetic energy bins. 
The first column reports the results for stably-trapped protons, concentrating in the SAA at PAMELA altitudes.
Constrained by the spacecraft orbit, the covered phase-space region varies with the magnetic latitude.
In particular, PAMELA can observe equatorial mirroring protons only for $L$-shell values up to $\sim$1.18 $R_{E}$, and
measured distributions 
are
strips of limited width parallel to the ``drift loss cone'', which delimits the $\alpha_{eq}$ range for which stable magnetic trapping does not occur.
Fluxes exhibit strong angular and radial de\-pen\-den\-cies.
PAMELA is able to measure trapped spectra up to their highest energies (about 4 GeV) \citep{PAMTRAPPED}.
For a comparison, Figure \ref{Fig1} also reports the fluxes for quasi- and un-trapped components. In this case,
measured maps\footnote{Note that the un-trapped flux suppression at highest energy and $L$ bins is due to the geomagnetic cut ($R<10/L^{3}$) used for selecting adiabatic trajectories.} result from the superposition of
distributions corresponding to regions characterized by a different local (or bounce) loss cone value. 
Fluxes are quite isotropic except in the SAA, where distributions are similar to those of stably-trapped protons \citep{ALBEDO}.

\begin{figure}[!t]
\centering
\includegraphics[width=3.5in]{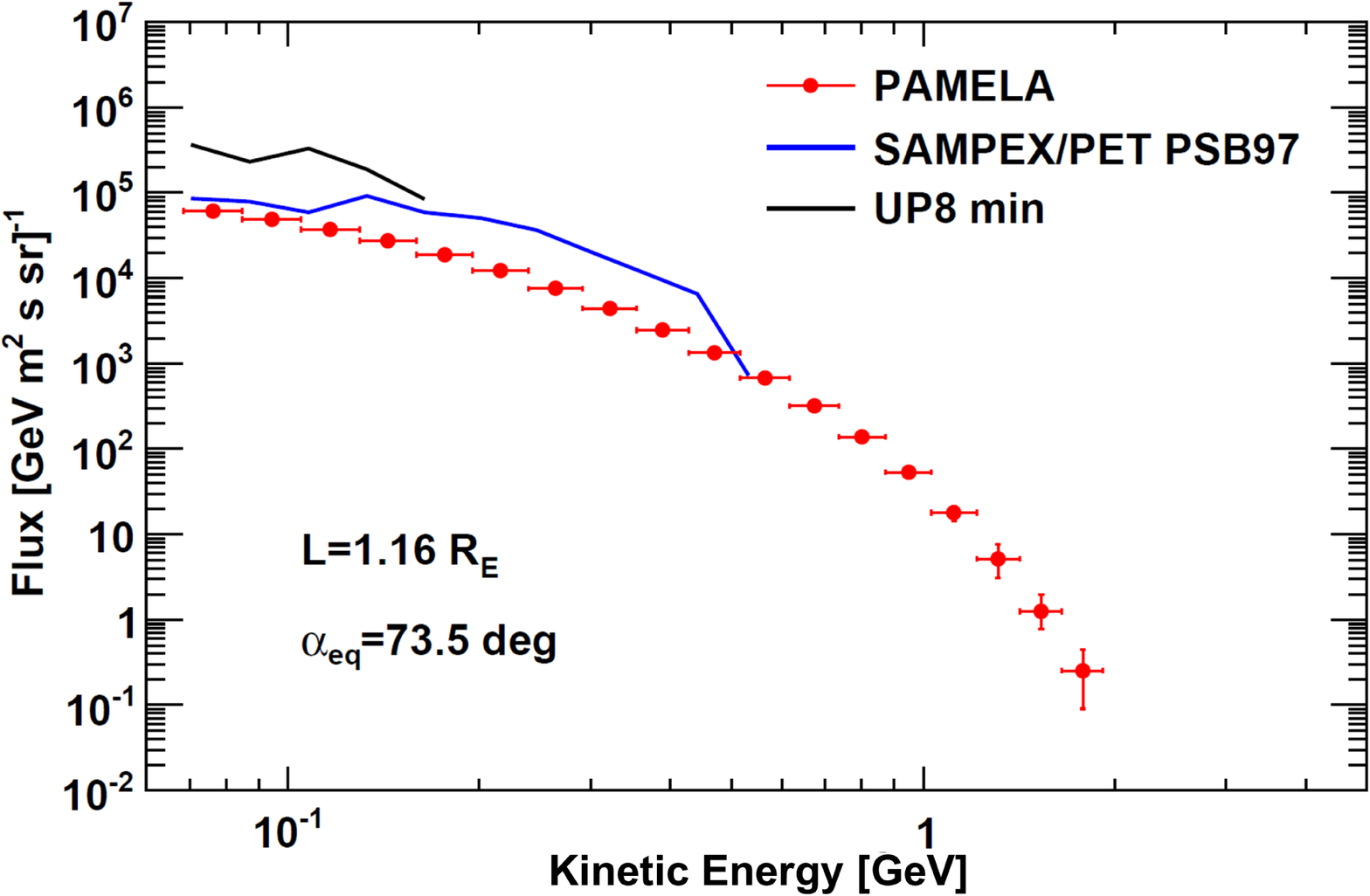}
\caption{PAMELA trapped proton energy spectrum for sample $\alpha_{eq}$ and $L$-shell values, compared with the predictions
from the UP8-min \citep{AP8,DALY} and the PSB97 \citep{PSB97} models (from SPENVIS \citep{SPENVIS}).
}
\label{Fig2}
\end{figure}

Figure \ref{Fig2} compares PAMELA geomagnetically trapped results with the predictions from two empirical models available in
the same energy and altitude ranges:
the AP8 \citep{AP8} unidirectional (or UP8 \citep{DALY}) model for solar minimum conditions, and the SAMPEX/PET PSB97 model \citep{PSB97}.
Data were derived by using the
SPENVIS web-tool \citep{SPENVIS}.
In general, the UP8 model significantly overestimates PAMELA observations, while a better agreement can be observed with the PSB97 model.
However,
PAMELA fluxes
do not show the spectral structures
present in the PSB97
predictions.  

\begin{figure}[!t]
\centering
\includegraphics[width=5.in]{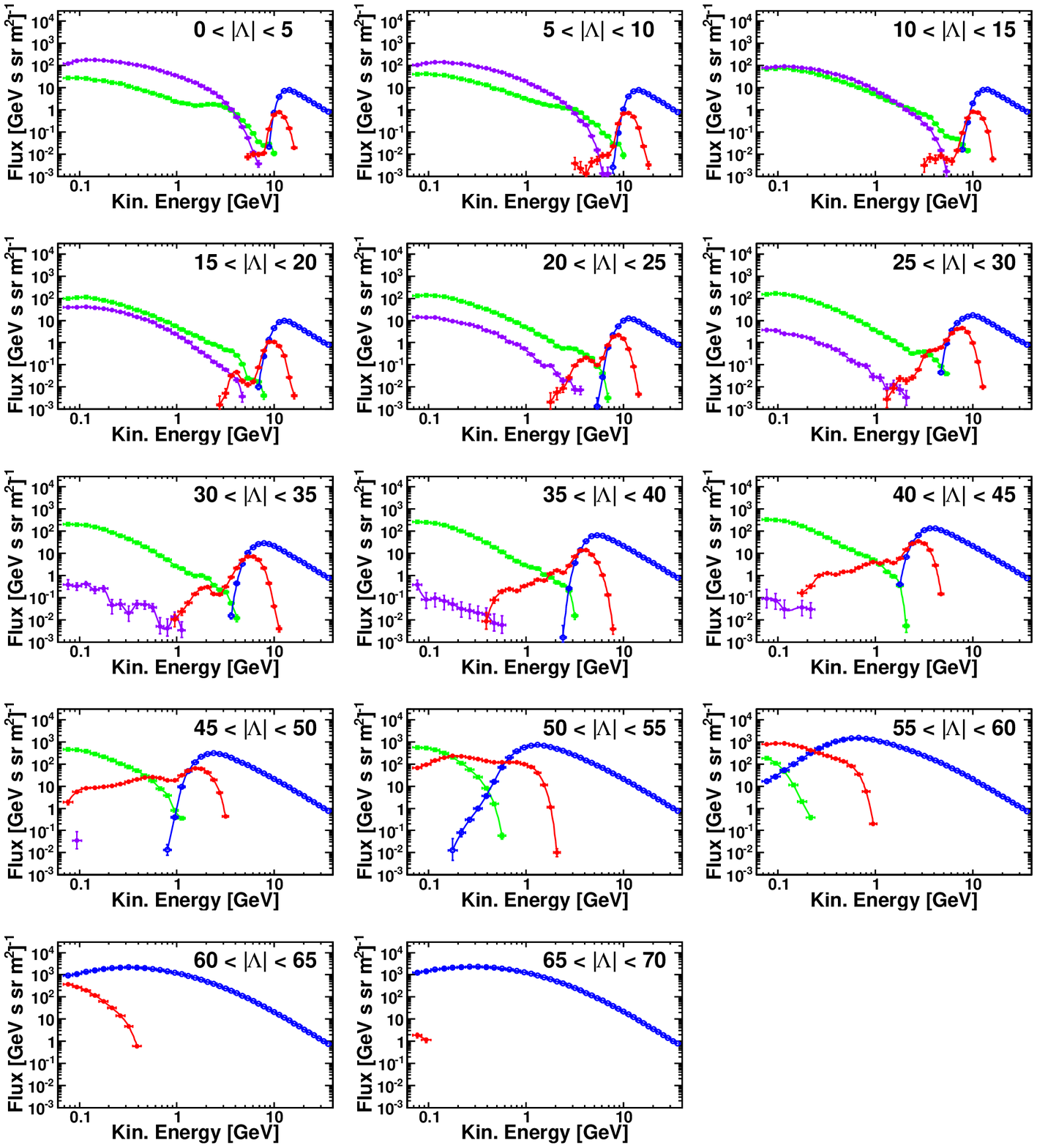}
\caption{Differential energy spectra (GeV$^{-1}$ m$^{-2}$ s$^{-1}$ sr$^{-1}$) outside the SAA for different AACGM latitude $|\Lambda|$ bins. Results for the several proton populations are shown: quasi-trapped (violet), precipitating (green), pseudo-trapped (red) and galactic (blue). Lines are to guide the eye.}
\label{Fig3}
\end{figure}
\begin{figure}[!t]
\centering
\includegraphics[width=5.1in]{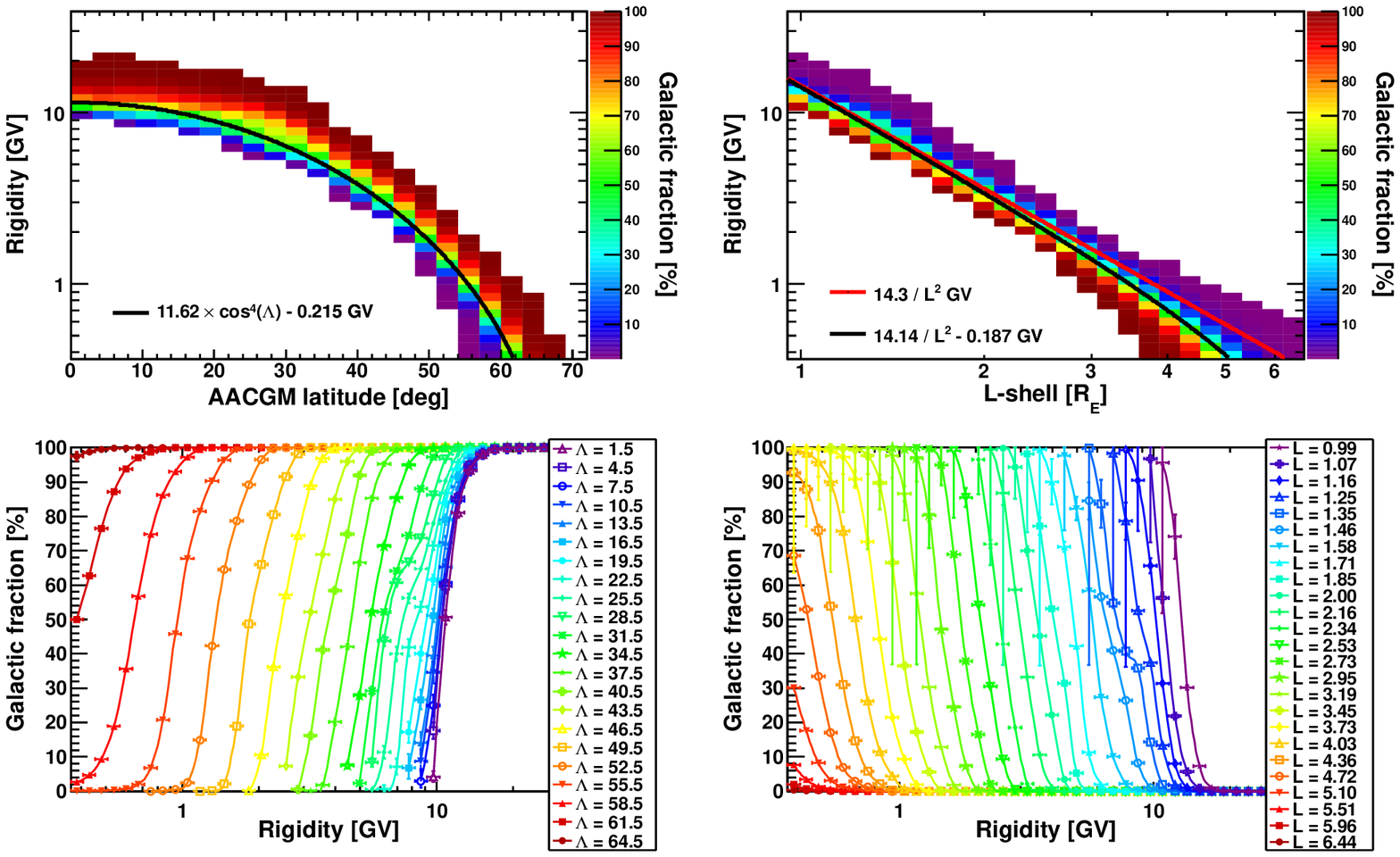}
\caption{Fraction of galactic protons in the penumbra region, as a function of particle rigidity and AACGM latitude $|\Lambda|$ (left) and McIlwain's $L$-shell (right). See the text for details.}
\label{Fig4}
\end{figure}

Albedo fluxes were mapped using the Altitude Adjusted Corrected Geomagnetic (AACGM) coordinates
\citep{Heres}, developed to provide a more realistic description of high latitude regions, by accounting for the multipolar geomagnetic field.
Figure \ref{Fig3} shows the spectra of the various albedo components outside the SAA (B$>$0.23 G) measured at different latitudes, 
along with the galactic component. Fluxes were averaged over longitudes. 
Quasi-trapped protons are limited to low latitudes and to energies below $\sim$ 8 GeV; their fluxes smoothly decrease with increasing latitude and energy.
Conversely, the precipitating component spreads to higher latitudes, with 
spectra extending up to $\sim$10 GeV.
Finally, 
pseudo-trapped protons concentrate at highest latitudes and energies (up to $\sim$ 20 GeV),
with a peak in the penumbra originated by large gyro-radius ($10^{2}\div10^{3}$ km) effects. 

Features of the penumbra region are investigated in Figure \ref{Fig4}, where the fraction of galactic over total (galactic + albedo) pro\-tons is displayed as a function of particle rigidity and AACGM latitude (left panels); for a comparison, di\-stri\-butions as a function of McIlwain's $L$-shell are also shown (right panels).
The penumbra was identified as the region where both albedo and galactic proton trajectories were reconstructed.
The black curves denote
a fit of points with an equal percentage of the two components, while the red line refers to the St\"{o}rmer vertical cutoff for the PAMELA epoch.
Bottom panels report 
cor\-responding rigidity profiles.

\section{Summary and Conclusions}\label{Conclusions}
PAMELA measurements of energetic ($>$70 MeV) under-cutoff proton fluxes at low Earth orbits (350$\div$610 km) have been presented. The detected proton sample 
was classified into geomagnetically trapped and re-entrant albedo on the basis of accurate particle tracing techniques.

\emph{Stably-trapped} protons, confined in the SAA at PAMELA altitudes, were investigated in the framework of the adiabatic theory.
PAMELA data extend the observational range for the trapped radiation down to lower $L$-shells ($\sim$ 1.1 $R_{E}$) and up to highest kinetic energies ($\lesssim$ 4 GeV), si\-gni\-fi\-cantly improving the description of the low altitude radiation environment,
where current models suffer from the largest uncertainties.

Albedo protons were classified into \emph{quasi-trapped} and \emph{un-trapped}: the former consist of re\-la\-ti\-ve\-ly long-lived protons populating the equatorial region, with trajectories similar to those of stably-trapped; 
the latter
include a short-lived (\emph{precipitating}) component spreading over all explored latitudes, along with a long-lived (\emph{pseudo-trapped}) component
concentrating near the geomagnetic cutoff and characterized by a chaotic motion (non-adiabatic trajectories).

PAMELA results significantly enhance the characterization of high energy proton populations
in a wide geomagnetic region,
enabling a more precise and complete view of atmospheric and magnetospheric effects
on the CR transport
near the Earth.


\begin{thebibliography}{99}

\bibitem[Adriani et al.(2013)]{SOLARMOD} O. Adriani, et al., 2013, %``Time dependence of the proton flux measured by PAMELA during the 2006 July -- 2009 December solar minimum'',
    ApJ 765:91.05205.

\bibitem[Adriani et al.(2014)]{PHYSICSREPORTS} O. Adriani, et al., 2014, Physics Reports, Vol. 544, 4, pp. 323--370, doi:10.1016/j.physrep.2014.06.003
%\bibitem{PICOZZA} P. Picozza, et al., 2007, Astropart. Phys., Vol 27, Pages: pp. 296-315.

\bibitem[Adriani et al.(2015a)]{PAMTRAPPED} O. Adriani, et al., 2015a, %``Trapped proton fluxes at low Earth orbits measured by the PAMELA experiment'',
ApJ 799 L4, doi:10.1088/2041-8205/799/1/L4.

\bibitem[Adriani et al.(2015b)]{ALBEDO} O. Adriani, et al., 2015b, %Reentrant albedo proton fluxes measured by the PAMELA experiment,
    J. Geophys. Res. Space Physics, 120, doi:10.1002/2015JA021019.

\bibitem[Alcaraz et al.(2000)]{AMS01b} J. Alcaraz, et al., 2000, Phys. Lett. B 472:215--226, doi:10.1016/S0370-2693(99)01427-6.
\bibitem[Bidoli et al.(2003)]{Bidoli} V. Bidoli, et al., 2003, J. Geophys. Res., 108, A5, 1211, doi:l0.1029/2002JA009684.
\bibitem[Cooke et al.(1991)]{Cooke} D. J. Cooke, Humble, J. E., Smart, M. A., et al., 1991, Il Nuovo Cimento, 14C, 213.


\bibitem[Daly \& Evans(1996)]{DALY} E. J. Daly \& H. D. R. Evans, 1996, Rad. Meas., vol. 26, no. 3, pp. 363-368.

\bibitem[Fiandrini et al.(2004)]{AMS01} E. Fiandrini, et al., 2004, J. Geophys. Res., 109, 10214, doi:10.1029/2004JA010394.
\bibitem[Finlay et al.(2010)]{IGRF11} C. C. Finlay, et al., 2010, %``International Geomagnetic Reference Field: the eleventh generation'',
    Geophysical Journal International, 183: 1216--1230.

\bibitem[Gussenhoven et al.(1993)]{CRRES} M. S. Gussenhoven, et al., 1993, IEEE Trans. Nucl. Sci., 40, 1450.
\bibitem[Gussenhoven et al.(1995)]{CRRES2} M. S. Gussenhoven, et al., 1995, IEEE Trans. Nucl. Sci., 42, 2035.


\bibitem[Heres \& Bonito(2007)]{Heres} W. Heres \& N. A. Bonito, 2007, Scientific Report AFRL--RV--HA--TR--2007--1190.
\bibitem[Heynderickx et al.(2002)]{SPENVIS} D. Heynderickx, et al., 2002, SAE Technical Paper 2000-01-2415.
\bibitem[Heynderickx et al.(1999)]{PSB97} D. Heynderickx, et al., 1999, IEEE Trans. Nucl. Sci., 46, 1475.
\bibitem[Hovestadt et al.(1972)]{Hovestadt} D. Hovestadt, et al., 1972, Phys. Rev. Lett. 28, 1340.
\bibitem[Huston et al.(1996)]{Huston96} S. L. Huston, G. A. Kuck and K. A. Pfitzer, 1996, Geophys. Monogr. Ser., Vol. 97, pp. 119--122.
\bibitem[Huston \& Pfitzer(1998)]{Huston} S. L. Huston \& K. A. Pfitzer, 1998, NASA Contract. Rep. NASA/CR-1998-208593.

\bibitem[Looper et al.(1996)]{Looper96} M. D. Looper, et al., 1996, Radiat. Meas., 26(6):967-78.
\bibitem[Looper et al.(1998)]{Looper98} M. D. Looper, J. B. Blake and R. A. Mewaldt, 1998, Adv. Space Res., 21(12):1679-82.

\bibitem[Meffert \& Gussenhoven(1994)]{CRRESPRO} J. D. Meffert \& M. S. Gussenhoven, 1994, PL-TR-94-2218, Environ. Res. Pap. 1158, Air Force Res. Lab., Wright-Patterson Air Force Base, Ohio.

\bibitem[Moritz(1972)]{Moritz} J. Moritz, 1972, Z. Geophys. 38, 701.

\bibitem[Sawyer \& Vette(1976)]{AP8} D. M. Sawyer \& J. I. Vette, 1976, NSSDC/WDC-A-R\&S 76-06.

\bibitem[Smart \& Shea(2000)]{TJPROG} D. F. Smart \& M. A. Shea, 2000, Final Report, Grant NAG5--8009, Center for Space Plasmas and Aeronomic Research, The University of Alabama in Huntsville.
\bibitem[Smart \& Shea(2005)]{SMART} D. F. Smart \& M. A. Shea, 2005, Adv. Space Res., 36, 2012--2020.

\bibitem[Sullivan(1971)]{Sullivan} J. D. Sullivan, 1971, Nucl. Instr. and Meth. 95, 5.

\bibitem[Treiman(1953)]{Treiman} S. B. Treiman, 1953, The Cosmic-Ray Albedo, Phys. Rev. 53, 957, doi:http://dx.doi.org/10.1103/PhysRev.91.957.

\bibitem[Tsyganenko \& Sitnov(2005)]{TS05} N. A. Tsyganenko \& M. I. Sitnov, 2005, J. Geophys. Res., Vol. 110, A03208.

\bibitem[Xapsos, et al.(2002)]{TPM1} M. A. Xapsos, et al., 2002, IEEE Trans. Nucl. Sci., 49, 2776.

\bibitem[Walt(1994)]{Walt} M. Walt, 1994, Introduction to Geomagnetically Trapped Radiation, Cambridge Univ. Press. (New York).


\end{thebibliography}
\end{document}